\documentclass[aps,twocolumn, prb,showpacs,groupedaddress,floatfix,longbibliography]{revtex4-1}

\usepackage{amsmath}
\usepackage{fixmath}
\usepackage{graphicx}
\usepackage[usenames,dvipsnames]{color}
\pdfoutput=1

\usepackage{color}
\usepackage{url}
\usepackage[colorlinks]{hyperref}
\hypersetup{%
	plainpages=true,
	breaklinks=true,
	hypertexnames=false,
	pageanchor=true,
	colorlinks=true,
	linkcolor={blue},
	citecolor={blue},
	urlcolor={blue}, 
	anchorcolor={black}
}

\newcommand{\figref}[1]{\mbox{Fig.~\ref{#1}}}

\newcommand{\secref}[1]{\mbox{Sec.~\ref{#1}}}

\newcommand{\appref}[1]{\mbox{Appendix~\ref{#1}}}
\renewcommand{\eqref}[1]{\mbox{Eq.~(\ref{#1})}}

\newcommand{\ket}[1]{\left|#1\right\rangle}
\newcommand{\bra}[1]{\left\langle#1\right|}

\newcommand{\sm}{\hat{\sigma}^-}
\renewcommand{\sp}{\hat{\sigma}^+}

\newcommand{\dfk}{\hat{d}_{F,k}^{}}
\newcommand{\dfck}{\hat{d}_{F,k}^\dag}

\newcommand{\dbk}{\hat{d}_{B,k}^{}}
\newcommand{\dbck}{\hat{d}_{B,k}^\dag}

\newcommand{\dpk}{\hat{d}_{\tilde{B},k}^{}}
\newcommand{\dpck}{\hat{d}_{\tilde{B},k}^\dag}
\newcommand{\dpmk}{\hat{d}_{\tilde{B},-k}^{}}
\newcommand{\dpcmk}{\hat{d}_{\tilde{B},-k}^\dag}

\newcommand{\bk}[1]{\beta_{\tilde{B},k}^{(#1)}}	

\newcommand{\be}{\begin{equation}}
	\newcommand{\ee}{\end{equation}}
\newcommand{\bea}{\begin{eqnarray}}
	\newcommand{\eea}{\end{eqnarray}}

\begin{document}
	
	
	\title{Polariton spectrum of the Dicke-Ising model}
	
	
	\author{Erika Cortese}
	\author{Luigi Garziano}
	\author{Simone De Liberato} 
	
	\affiliation{School of Physics and Astronomy, University of Southampton, Southampton, SO17 1BJ, United Kingdom}
		
	\begin{abstract}
	The Dicke-Ising model describes cavity quantum electrodynamics setups in which dipoles couple not only with the photonic cavity field but also to each other through dipole-dipole interaction. In this work we diagonalise such a model in terms of bosonic polaritonic operators for arbitrarily large values of the light-matter coupling and for values of the dipole-dipole interaction until the onset of the ferromagnetic Ising phase transition. 
	In order to accomplish this task we exploit higher order terms of the Holstein-Primakoff transformation, developing a general method which allows to solve the normal phase of the Ising model in term of bosonic excitations for large values of the dipole-dipole interaction. Our results shed light on the interplay between the dipole-dipole and the light-matter coupling strengths, and their effect on the virtual excitations which populate the ground-state when the interactions become comparable with the bare transition frequency.	
	\end{abstract}
	
	
	\maketitle
	

\section{Introduction}

The strong coupling regime of cavity quantum electrodynamics (QED) where cavity photons interact with matter at a rate exceeding relaxation and dephasing has been a main focus of research, leading to a wide number of scientific and technological groundbreaking results \cite{Haroche,Kavokin,Agranovich}.
While achieving strong coupling for single emitters is an extremely challenging feat \cite{Haroche13,Baumberg16}, such a regime has been reached in numerous solid state systems in which many dipoles can coherently interact to the same photonic mode. In this case in fact the dipoles collectively interact to the electromagnetic field, enhancing the coupling strength by the square root of their number.  The Dicke model describing such systems \cite{Dicke54} can be  solved exactly in the diluite linear regime treating the matter excitations as bosons and solving the resulting Hamiltonian using a multimode Bogoliubov transformation. This procedure, originally used by Hopfield \cite{Hopfield58}, leads to an energy spectrum composed of bosonic hybrid light-matter quasi-particles called polaritons.
When a non-negligible longitudinal coupling between the different dipoles is present, like in the well-studied case of $ J $-aggregates, the matter Hamiltonian describing the dipoles and their mutual interaction takes the form of an Ising model in a transverse field. In particular, the one-dimensional Ising Hamiltonian, which is relevant for the quantum description of one-dimensional exciton systems such as molecular aggregates \cite{Agranovich,Mukamel89}, can be solved exactly in terms of fermionic excitations through the well-known Jordan-Wigner transformation \cite{Jordan28,Fradkin89,Bakalis97}.
While at least in certain regimes exact approaches exist to solve both the Dicke and the Ising models separately, those methods do not work well together and as a consequence the full Dicke-Ising model, which includes both the longitudinal dipole-dipole interaction and the coupling with the transverse cavity photons, has until now been considered only from a thermodynamical perspective. Specifically, it has been studied both analytically by using a mean field approach to investigate the model's first- and second-order phase transitions \cite{Gammelmark11,Zhang14a}, and numerically by using matrix product states \cite{Gammelmark12}. Recently, more complicated Dicke-Ising systems  with variable Ising interactions and position-dependent coupling to a resonator field have also been considered \cite{Zhang14b}. 

The interest for the theoretical study of this kind of systems has been motivated not only by the experimental observation of organic molecular systems displaying very large values of light-matter coupling strength  \cite{Lidzey98,Coles14a,Coles14b,Dintinger05,Hakala09,Gambino14,Gubbin16}, but also by recent advances in the field of circuit QED, in which arrays of thousands of superconducting qubits coupled to a single photonic resonator have been realised \cite{Kakuyanagi16}. In those systems large values of both the light-matter coupling, and of the inter-dipole coupling can be achieved \cite{Majer07,Niemczyk10,Forn10,Koch10,You11,Yoshihara17,Roy17}. These systems may provide an excellent platform for quantum technologies, including the development of quantum simulations \cite{Schmidt13,Georgescu14} and quantum metamaterials \cite{Zheludev10,Soukoulis11}.

Cavity QED systems displaying the so-called ultrastrong coupling regime, in which the light-matter coupling becomes comparable to the typical frequencies of the system, have been arousing growing interest in the last decade. One of the main reasons is that  they exhibit a rich new phenomenology \cite{Werlang08,Cao11,Ridolfo12,Stassi13,DeLiberato14,Garziano16} largely ascribed to the presence of a finite population of virtual excitations in the ground-state \cite{Ciuti05,DeLiberato07,DeLiberato09,Garziano13,Law13,DeLiberato17}. This effect is essentially due to the fact that in this regime the routinely-invoked rotating wave approximation (RWA) breaks down and the counter-rotating terms in the interaction Hamiltonian have to be explicitly taken into account.

In this article, we will perform an analytic diagonalisation of the Dicke-Ising Hamiltonian. In order to accomplish this task, we will start by developing an approach able to correctly diagonalise the Ising model for a one-dimensional array of $N$ interacting dipoles in terms of bosonic modes by considering higher orders in the Holstein-Primakoff transformation \cite{Holstein40}. After proving that the first non-trivial order already provides results exact to the second order in the dipole-dipole interaction for the Ising model, we will diagonalise the full Dicke-Ising Hamiltonian in terms of bosonic polariton quasi-particles.
Our approach allows us to treat arbitrary values of the light-matter coupling, and the dipole-dipole interaction until the onset of the Ising ferromagnetic phase transition. As mentioned before, this implies that the number-conserving approximations which are often exploited to simplify both the Dicke and the Ising model in transverse field  under the respective names of RWA and Heitler-London approximation (HLA) \cite{Agranovich, Bakalis97, Agranovich00}, cannot be applied since they break down when the relevant coupling parameter becomes a non-negligible fraction of the bare transition frequency.
Our results allow us to make predictions for  different kinds of physical implementations, demonstrating how the interplay between the two different kinds of coupling modifies the nature of the ground-state.

The outline of this article is as follows. In \secref{secA} we  start by introducing two approaches usually employed to solve the Ising model in transverse field, discussing their respective shortcomings. We will then introduce a novel approach which allows us to recover the correct spectrum of the Ising model in terms of bosonic excitations for arbitrary values of the dipole-dipole interaction. Exploiting the results obtained in  \secref{secA}, in \secref{secB} we will introduce and solve the Dicke-Ising model using different approaches of increasing sophistication.  For sake of clarity we kept in the main body of the paper only a minimum of formulas, moving more technical calculations into the Appendices.

\section{The Ising Hamiltonian} \label{secA}
\subsection{Definitions and background} \label{secA1}
\begin{figure}[!h]
	\centering
	\includegraphics[scale=0.5
	]{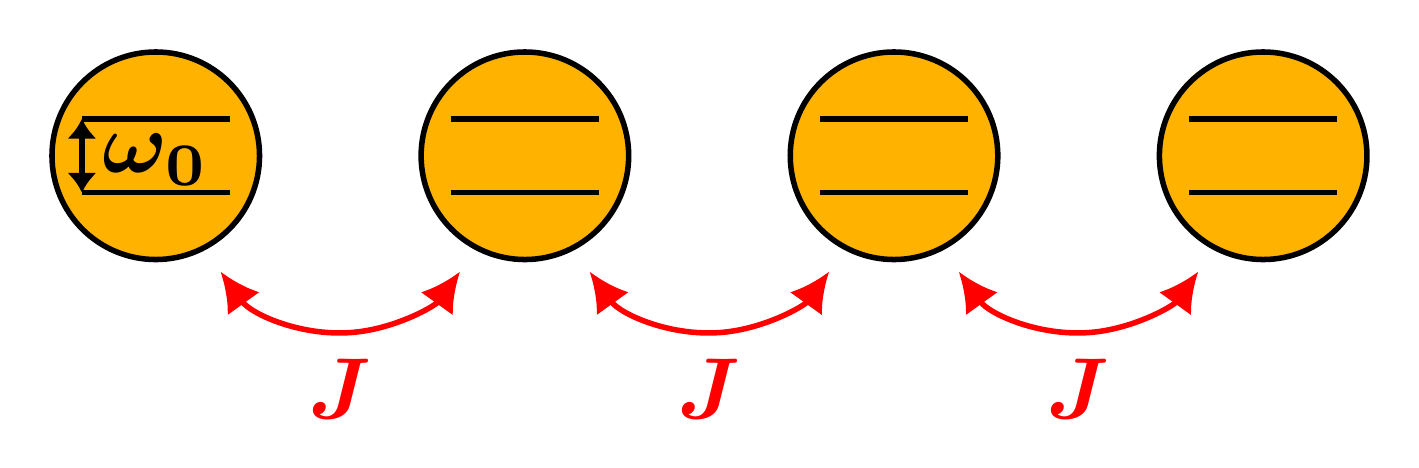}
	\includegraphics[scale=0.38
	]{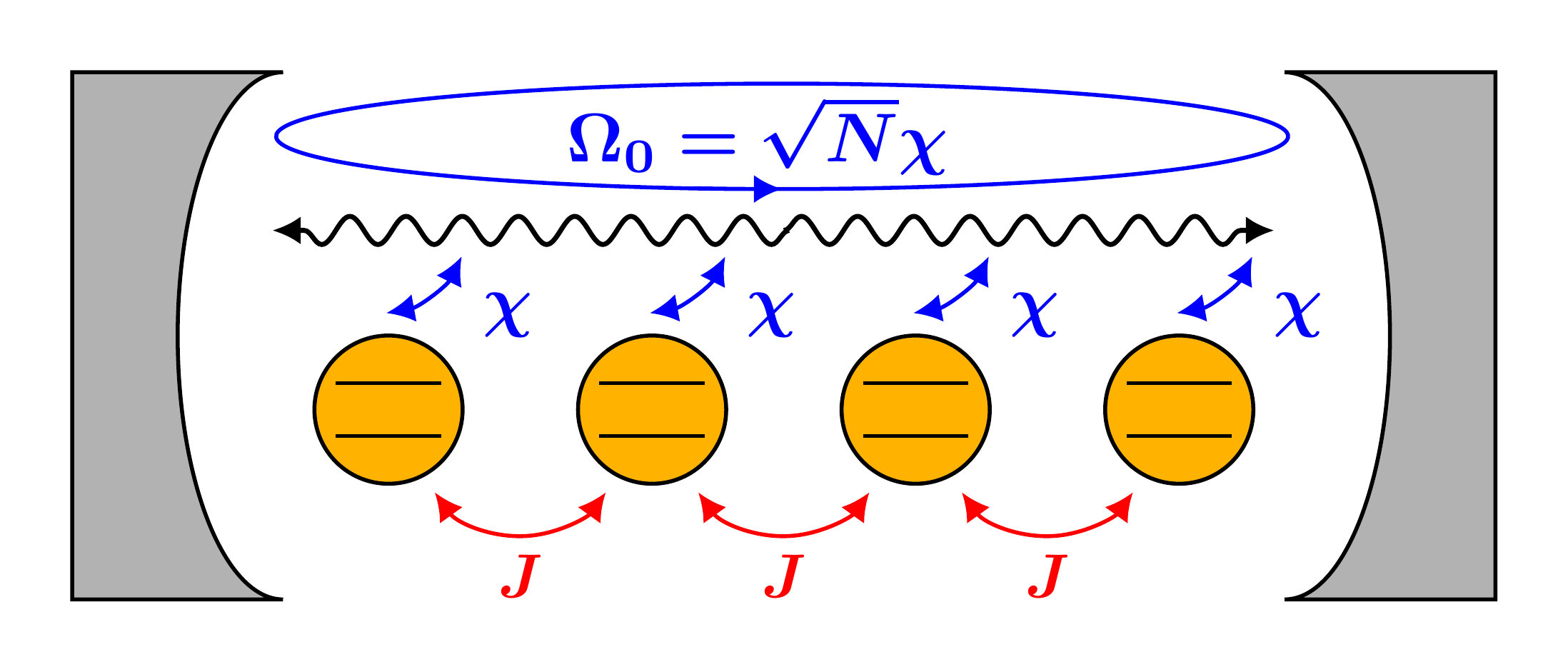}
	\caption{Top: One dimensional chain of $ N $ coupled two level systems with transition frequency $\omega_0$ and next-neighbour coupling strength $J$. Bottom: The same chain of the top panel coupled to a cavity photonic field through a single-dipole light-matter coupling $\chi$ and collective coupling $\Omega_0=\sqrt{N} \chi$.    }\label{fig:11}
\end{figure}
A one-dimensional chain of $N$ coupled two-level systems, sketched in the top panel of \figref{fig:11}, can be described by the Hamiltonian (throughout the paper, we set $\hbar=1$):
\be \label{1}
\hat{H}_{ M}=\sum_{n=1}^{N}\omega_{0}\,\sp_{n} \sm_{n}+J \sum_{n=1}^{N-1}(\sp_{n} +\sm_{n})(\sp_{n+1}+\sm_{n+1}),
\ee 
where $ \omega_{0} $ and $J$ are, respectively, the dipole transition frequency and the next-neighbour dipole-dipole coupling strength (Ising coupling), and the $n$th dipole is described by a two level system with Pauli rising and lowering operators $\hat{\sigma}_{n}^{+}$ and  $\hat{\sigma}_{n}^{-}$.  As detailed in \appref{appA1}, exploiting the Jordan-Wigner transformation			
the Hamiltonian in \eqref{1} can be put in the diagonal form   
\be \label{7}
\hat{H}_{ F}=\sum_{k}E_{F,k}\, \dfck \dfk +E_{F}^{0},
\ee
as function of the operators $\dfck,\dfk$, obeying fermionic anticommutation relations  
\be
\Bigl\{\hat{d}_{F,k}^{},\hat{d}_{F,k'}^\dag\Bigr\}=\delta_{k k'}\hspace{0.1cm},\hspace{0.1cm}  \Bigl\{\hat{d}_{F,k}^\dag,\hat{d}_{F,k'}^\dag\Bigr\}=\Bigl\{\hat{d}_{F,k},\hat{d}_{F,k'}\Bigr\}=0,
\ee
where $k$ is the quasi-momentum $0 < k < \pi$ 
The single-particle energies $E_{F,k}$ are 
\be 
E_{F,k}= (\omega_{0}^2+4J^2+4J\omega_{0}\cos k)^{1/2},  \label{12}
\ee
and  the ground-state energy takes the form
\be 
\label{12bis}
E_{F}^{0}=\frac{1}{2}\Biggl(N\omega_{0}-\sum_{k}E_{F,k}\Biggr).
\ee
Note that from \eqref{12} we see that the spectrum is invariant under the transformation $J \rightarrow -J $, $k\rightarrow \pi-k$, 
and the system presents the well-known ferromagnetic phase transition for the normalised Ising coupling $ \eta\equiv J/\omega_0=0.25$. As we are only interested in the normal phase, in the rest of the paper we will thus limit ourself to consider $\lvert \eta \lvert \leq 0.25$.

Unfortunately, such an exact solution is inadequate to rigorously determine the polaritonic spectrum when the Ising chain is strongly coupled to a cavity photonic field. Indeed, the first issue is that one would then end up with a coupled light-matter Hamiltonian including both fermionic and bosonic operators, and thus not adapted to be solved by standard multimode Bogoliubov transformation. Moreover, the Jordan-Wigner mapping leads to a Hamiltonian which depends upon the parity of the excitation number, a quantity which is not conserved when the coupling with a cavity is taken into account.

A simpler and widespread approach consists of mapping the Hamiltonian in \eqref{1} into a fully bosonic one through the Holstein-Primakoff transformation
\begin{align} \nonumber
\hat{\sigma}_{n}^{+}=& \,\hat{b}^{\dag}_{n}\sqrt{1-\hat{b}^{\dag}_{n}\hat{b}_{n}}=\hat{b}^{\dag}_{n}\Biggl(1-\frac{\hat{b}^{\dag}_{n}\hat{b}_{n}}{2}-\frac{\hat{b}^{\dag}_{n}\hat{b}_{n}\hat{b}^{\dag}_{n}\hat{b}_{n}}{8}-\dots \Biggr),\\\label{14}
\hat{\sigma}_{n}^{-}=&\, \sqrt{1-\hat{b}^{\dag}_{n}\hat{b}_{n}}\,\hat{b}_{n}=\Biggl(1-\frac{\hat{b}^{\dag}_{n}\hat{b}_{n}}{2}-\frac{\hat{b}^{\dag}_{n}\hat{b}_{n}\hat{b}^{\dag}_{n}\hat{b}_{n}}{8}-\dots\Biggr) \,\hat{b}_{n}, 
\end{align}
where the $\hat{b}^{\dag}_{n}$ operators obey bosonic commutation relations
\be
\bigl[\hat{b}_{n},\hat{b}_{n'}^\dag\bigr]=\delta_{n n'}\hspace{0.1cm},\hspace{0.1cm}  \bigl[\hat{b}_{n}^\dag,\hat{b}_{n'}^\dag\bigr]=\bigl[\hat{b}_{n},\hat{b}_{n'}\bigr]=0.
\ee
If one assumes the population of the $n$th dipole $\hat{b}^{\dag}_{n}\hat{b}_{n}$  to be negligible, only the first term in the expansions of \eqref{14} can be retained, an approximation which, following the literature, we will refer to as the Bose approximation.
In this case, the Hamiltonian of \eqref{1}  becomes:
\begin{align} \label{bos}
\hat{H}_{\rm B}=\,& \omega_{\rm 0} \sum_{n=1}^{N}\hat{b}_{n}^\dag \hat{b}_{n}+J\sum_{n=1}^{N-1}(
\hat{b}_{n}+\hat{b}_{n}^\dag)( \hat{b}_{n+1}+\hat{b}_{n+1}^\dag).\\\nonumber
\end{align}
As detailed in \appref{appA2}, following a procedure similar to the one used in the exact diagonalisation, one can write 
the Hamiltonian in the diagonal form: 
\be \label{25}
\hat{H}_{B}=\sum_{k}E_{B,k}\, \dbck \dbk+E_{B}^{0},
\ee
where $ \dbck,\dbk $ are bosonic operators obeying the commutation rules 
\be \bigl[ \hat{d}_{B,k}^{},\hat{d}_{B,k'}^\dag \bigr]=\delta_{k k'}\hspace{0.1cm},\hspace{0.1cm}  \bigl[\hat{d}_{B,k}^\dag,\hat{d}_{B,k'}^\dag \bigr]=\bigl[ \hat{d}_{B,k},\hat{d}_{B,k'} \bigr]=0, 
\ee and
\bea \nonumber
E_{B,k}&=&(\omega_{0}^2+4J\omega_{0}\cos k)^{1/2},\\ \label{26} E_{B}^{0}&=&\frac{1}{2}\Biggl(\sum_{k}E_{B,k}-N\omega_{0}\Biggr),
\eea
are, respectively, the single-particle energy and the ground-state energy. Note that from here on we will use Pekar boundary conditions \cite{Pekar1957} for the wavevectors 
\be
\label{Pekar} 
k\equiv k(l)=l \pi/(N+1),   
\ee
with $ l=1,2,\dots ,N$.

In order to make a direct comparison between the exact theory and the Bose approximation, it is useful to consider the expansions up to the second order in $\eta$  for the single-particle energies $ E_{F,k} $ and $ E_{B,k} $ from Eqs. (\ref{12}) 
and (\ref{26}). These expansions, which are given by
\begin{align}
\frac{E_{F,k}}{\omega_0}\approx\,& 1+ 2\cos k \,\eta+2\sin^2 k \,\eta^2,\nonumber\\\label{46}
\frac{E_{B,k}}{\omega_0}\approx\,& 1+ 2\cos k \,\eta-2\cos^2 k \,\eta^2,
\end{align}
show that the Bose approximation breaks down already at the second order in $\eta$.

\subsection{Virtual excitations in the Ising model} \label{secA2}
Being the Holstein-Primakoff mapping exact, the discrepancy between \eqref{12} and \eqref{26} can only be attributed to the truncation of the series in  \eqref{14}. This in turn implies that our assumption of a negligible excitation density in the ground-state of the Ising model is fallacious for non-negligible values of the coupling strength $\eta$. The failure of the Bose approximation to yield the correct results can be explained by considering that, as shown in \appref{appA2}, the ground-state of the Bose approximation contains a finite population of virtual excitations. 
That is, calling $\ket{G}$ the ground-state of \eqref{bos}, the virtual ground-state population on an arbitrary site
can be written as 
\be\label{28}
\mathcal{N}\equiv \bra{G}\hat{b}^{\dag}_{n}\hat{b}_{n}\ket{G}= \frac{\eta^2}{2}+O(\eta^3),
\ee   
which is non-vanishing in the thermodynamic limit $N\rightarrow\infty$.
When such a quantity is non-negligible, a development of  \eqref{14} to the lowest order becomes thus inaccurate. Note that, on the contrary, in the Dicke model the ground-state virtual excitation density goes to zero in the thermodynamic limit, and the lowest order expansion of the Holstein-Primakoff transformation is in this case exact for the single-particle states.

\subsection{First-order Holstein-Primakoff approximation: results and discussion} \label{secA3}

In order to overcome the difficulties considered in the previous Section and diagonalise \eqref{1} in terms of bosonic quasi-particles for sizeable values of $\eta$, here we introduce an approach to recover a quadratic bosonic Hamiltonians from the higher-order terms of the Holstein-Primakoff expansion. When higher order terms from \eqref{14} are considered, the Hamiltonian will generally contain nonlinear terms composed of an even number of $\hat{b}_n$ and $\hat{b}^{\dagger}_n$ operators.
In analogy with what done for the Bose approximation in \appref{appA2}, we aim to put such an Hamiltonian in diagonal form introducing the bosonic operators $\dpk$ defined by the relations
\begin{align} \nonumber 
\hat{b}_{n}=& \frac{1}{\sqrt{N}}\sum_{k} e^{ink}(\alpha_{\tilde{B},k }\,\dpk+\beta_{\tilde{B},k}\,\dpcmk),\\\label{34}
\hat{b}^{\dag}_{n}=& \frac{1}{\sqrt{N}}\sum_{k} e^{-ink}(\beta_{\tilde{B},k}\,\dpmk+\alpha_{\tilde{B},k }\,\dpck), 
\end{align} 
where the Bogoliubov coefficients $\alpha_{\tilde{B},k}$ and $\beta_{\tilde{B},k}$,
which can always be chosen as real and must satisfy the canonical relation $ \alpha_{\tilde{B},k }^2-\beta_{\tilde{B},k}^2=1$, are determined diagonalising the normally-ordered quadratic part of the nonlinear Hamiltonian. If $\ket{\tilde{G}}$ is the ground-state of such a quadratic part, defined by the relations $\dpk \ket{\tilde{G}}=0$, then all remaining nonlinear terms annihilate both $\ket{\tilde{G}}$ and all the single-particle states $\dpck \ket{\tilde{G}}$. This means that \eqref{34} exactly diagonalises the system to the chosen order in the single-particle sector. Note that the price to pay for this procedure is that the equations which determine the Bogoliubov coefficients are now nonlinear, and their degree and complexity increase with the order of the nonlinear terms retained in \eqref{14}.
While such a nonlinear set of equations could be solved numerically, in the following we will instead adopt a perturbative approach to order $\eta^2$. Not only this is consistent with the fact that, from \eqref{28}, we can interpret \eqref{14} as a series in $\eta^2$, but it will also allow us to derive analytic expressions which can be compared to those obtained by other approaches in the previous Sections. 

\begin{figure}[!ht]
	\centering
	\includegraphics[scale=0.6
	]{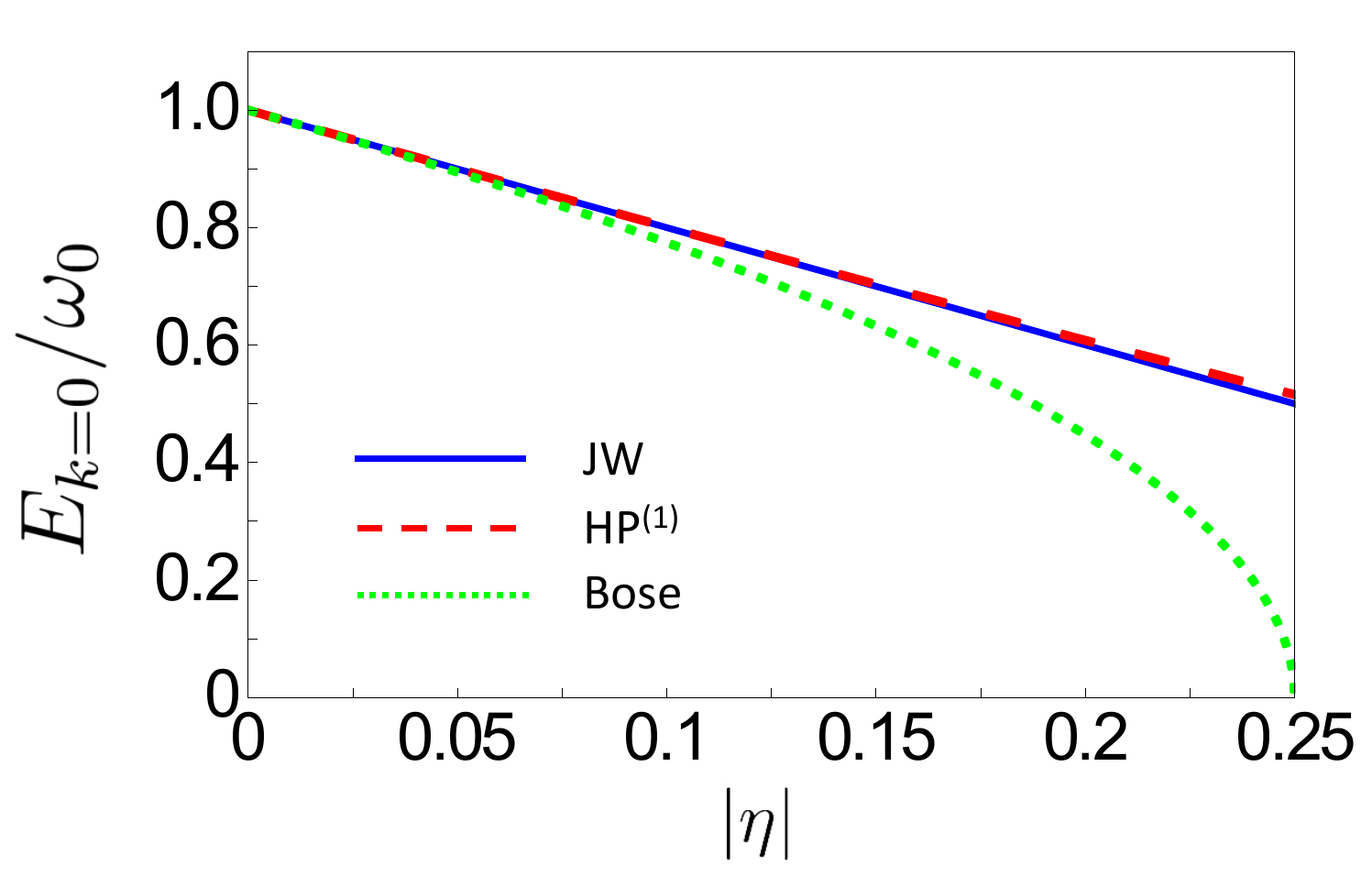}
	\caption{Single-particle energies for $ k=0 $ obtained as a function of the normalised Ising coupling strength  $ |\eta| $. Shown are the results obtained with the exact theory (solid blue line), the Bose approximation (dotted green  line) and the first-order Holstein-Primakoff approximation (dashed red line). It can be observed that, unlike the Bose approximation, the first-order Holstein-Primakoff approximation is in good agreement with the exact theory even for large values of $ |\eta| $.}
	\label{fig:1}
\end{figure}

Although the method outlined above is applicable to all orders, in the rest of this paper we will consider only the first nonlinear term in the expansion of \eqref{bos}, which allows to obtain quantitatively correct results also for $\eta$ close to the ferromagnetic phase transition boundary, 
that is we will consider
\begin{align} \nonumber
\hat{\sigma}_{n}^{+}=& \,\hat{b}^{\dag}_{n}\Biggl(1-\frac{\hat{b}^{\dag}_{n}\hat{b}_{n}}{2} \Biggr),\\\label{30}
\hat{\sigma}_{n}^{-}=&\,\Biggl(1-\frac{\hat{b}^{\dag}_{n}\hat{b}_{n}}{2}\Biggr) \,\hat{b}_{n}. 
\end{align} 
We will refer to \eqref{30} as first-order Holstein-Primakoff approximation, while keeping the name Bose approximation for what is effectively the zeroth-order approximation.
Using \eqref{30} the Hamiltonian in \eqref{1} can be written as
\be \label{32} 
\hat{H}_{ M}=\hat{H}_{ M}^{(0)} +\hat{H}_{ M}^{(1)},
\ee
where $ \hat{H}_{ M}^{(0)}$ is given by \eqref{bos} and the nonlinear part is
\be \label{33}
\hat{H}_{ M}^{(1)}=-\frac{J}{2}\,\hat{b}^{\dag}_{n}\Bigl[\Bigl(\hat{b}^{\dag}_{n}+\hat{b}_{n}\Bigr)\Bigl(\hat{b}^{\dag}_{n-1}+\hat{b}_{n-1}+\hat{b}^{\dag}_{n+1}+\hat{b}_{n+1}\Bigr) \Bigr]\hat{b}_{n}.
\ee
Using \eqref{34} in \eqref{32}, putting all the terms in normal order, and keeping only terms up to the second order, we obtain
\begin{align}\nonumber
\hat{H}_M'=&\sum_{k}\Bigl[\mathcal{A}_k-\frac{J}{2} \sum_{k'} f(k,k') \Bigr] \dpck \dpk\\\nonumber
+&\sum_{k}\Bigl[\mathcal{B}_k-\frac{J}{2} \sum_{k'} g(k,k')     \Bigr]\Bigl(\dpck \,\dpcmk +\dpk \,\dpmk\Bigr)\\\label{36}
+&\sum_{k}\Bigl[\mathcal{C}_k-\frac{J}{2} \sum_{k'} h(k,k') \Bigr],
\end{align}
where the expression of all the coefficients can be found in \appref{appB}.
Diagonalising the Hamiltonian is thus equivalent to solve the set of equations
\be
\mathcal{B}_k-\frac{J}{2} \sum_{k'} g(k,k')=0,\quad \forall k, \label{S1}
\ee 
which can be done analytically expanding the coefficients $ \alpha_{\tilde{B},k } $ and $ \beta_{\tilde{B},k} $ to the second order in $ \eta $.
Following such a procedure, detailed in  \appref{appB}, we obtain
\begin{align}\label{42}
\alpha_{\tilde{B},k }&\approx1+\frac{\cos^2 k}{2} \,\eta^2,\\ 
\beta_{\tilde{B},k }&\approx -\cos k\, \eta+\Bigl(2\cos^2 k-\frac{1}{2}\Bigr) \,\eta^2,\nonumber
\end{align}
resulting in single-particle energies 
\be \label{HP1en}
\frac{E_{\tilde{B},k}}{\omega_0}\approx 1+ 2\cos k \,\eta+2\sin^2 k \,\eta^2,
\ee
which, to the considered order, coincide with the exact result from \eqref{46}.
Figure~\ref{fig:1} displays the normalised single-particle energies $ E_{F,k}/\omega_0   $ (solid blue line), $ E_{B,k}/\omega_0 $ (dashed red line), and  $ E_{\tilde{B},k}/\omega_0   $ (dotted green  line) calculated as a function of $ |\eta| $ for the first excited mode, which, in the thermodynamic limit,   corresponds to $ k\rightarrow 0 $. It can be observed that, while for weak values of $ \eta $ the three curves overlap, when the coupling strength $ J $ becomes a significant fraction of $ \omega_0 $ the Bose approximation presents a significant deviation from the exact theory. On the contrary, the results obtained from the first-order Holstein-Primakoff approximation are in very good agreement with the exact theory.

\begin{figure}[!h]
	\centering
	\includegraphics[scale=0.5
	]{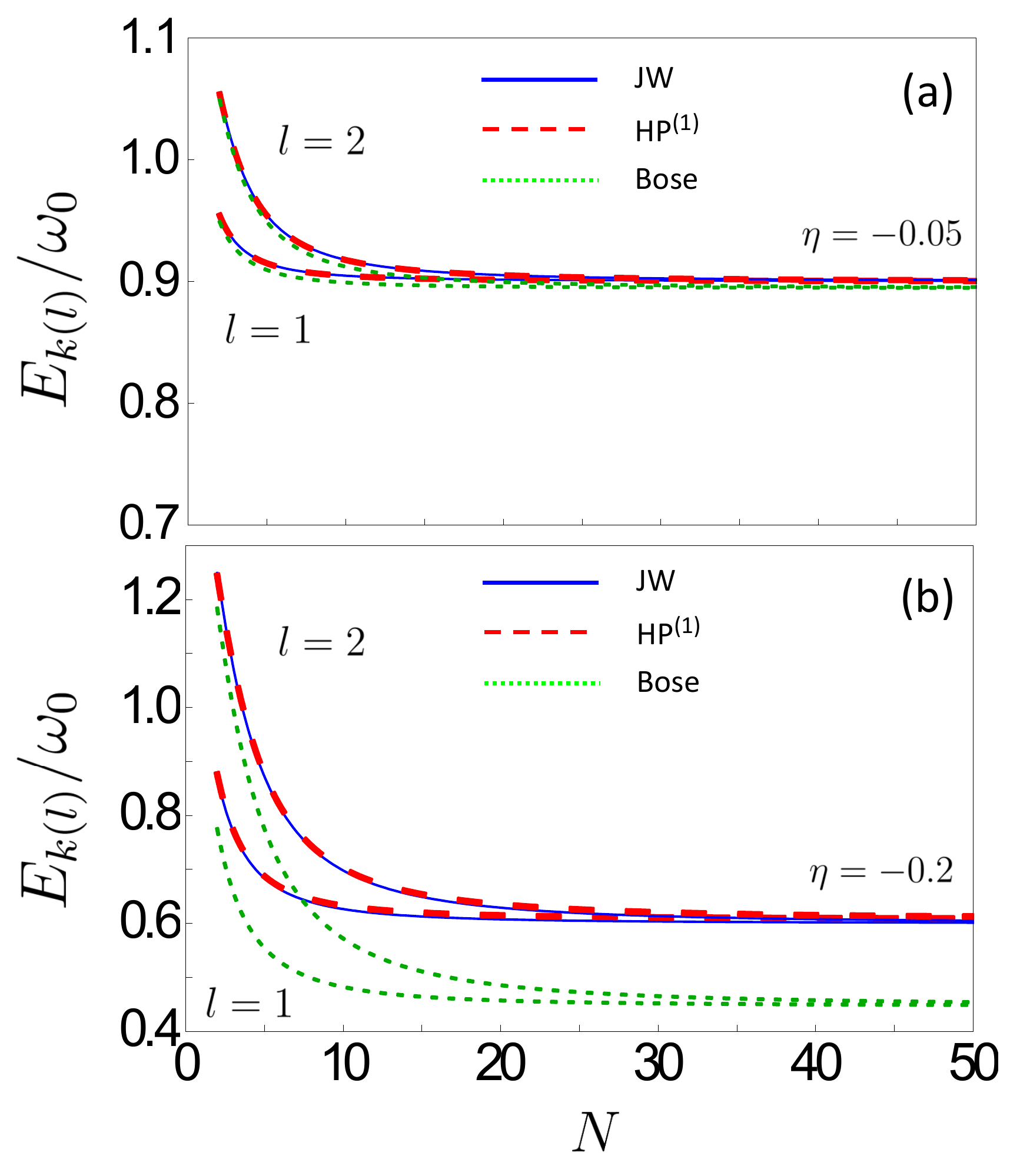}
	\caption{First two excited states  $E_{k(l=1)} $ and $E_{k(l=2)} $  as a function of the chain length $ N $ for two different values of the normalised coupling strength: (a) $ \eta = -0.05 $   and (b) $ \eta = -0.2 $. Shown are the results obtained with the exact theory (solid blue line), the Bose approximation (dotted green  line) and the first-order Holstein-Primakoff approximation (dashed red line).}
	\label{fig:2}
\end{figure}

We next turn to the study of the dependence of the single-particle energies on the number $ N $ of dipoles. 
In \figref{fig:2} we display the first two excited states (corresponding to $ l=1 $ and  $ l=2 $, where $l$ is the mode index from \eqref{Pekar})  as a function of the chain length $ N $ for (a) $ \eta = -0.05 $ and (b) $ \eta = -0.2 $. Once again, given are the results for the exact theory (\eqref{12}), the Bose approximation (\eqref{26}) and the first-order Holstein-Primakoff approximation (\eqref{HP1en}). It can be observed that the Bose approximation gives good results only for weak couplings, while the results obtained by applying the first-order Holstein-Primakoff approximation are robust also for large values $ |\eta| $.   

Finally, we can calculate the ground-state energy in the first-order Holstein-Primakoff approximation up to the second order in $ \eta $, 
\be \label{HPground}
\frac{E_{\tilde{B}}^{0}}{\omega_0}\approx \sum_{k}\cos k\Bigl[\Bigl(1-\cos k\Bigr)\eta^2 \Bigr]\approx\frac{\eta^2}{2}.
\ee 
A comparison between \eqref{HPground} and Eqs.~(\ref{12}) and (\ref{26}) shows that, to the second order in $ \eta $, the three approximations give the same results for the ground-state energy.

\section{Dicke-Ising Hamiltonian in the bosonic framework}\label{secB}

\subsection{The Dicke-Ising Hamiltonian}
In the present Section we study a one-dimensional Ising chain of $N$ dipoles, with lattice constant $a$, coupled to a photonic field in a resonant cavity of length $L=a(N+1)$, as described in the bottom panel of \figref{fig:11}.
Our main purposes will be to analytically determine the polariton spectrum of the coupled system and to investigate the interplay between the Ising and light-matter couplings.
The overall system can be described by the Dicke-Ising Hamiltonian
\be \label{H}
\hat{H}_{DI}=\hat{H}_{C}+\hat{H}_{M}+\hat{H}_{I}+\hat{H}_{D},
\ee
with
\begin{align}\label{sigmaLM}
\hat{H}_{C}&=\sum_{k} \omega_{k} \,\hat{a}_{k}^\dag\hat{a}_{k}, \\ \label{sigmaI}
\hat{H}_{I}&=\sum_{n=1}^{N} \sum_{k}  \chi \,(\hat{a}_{k}e^{ink}-\hat{a}_{k}^\dag e^{-ink})( \hat{\sigma}_{n}^--\hat{\sigma}_{n}^+),\\
\hat{H}_{D}&=\sum_{k} D_{k} (\hat{a}_{k}^\dag+\hat{a}_{-k})(\hat{a}_{-k}^\dag+\hat{a}_{k}),
\end{align}
where $\hat{H}_{C}$ represents the bare energy of the photonic modes of frequencies $\omega_k=c k/L$, with bosonic annihilation and creation operators $\hat{a}_{k}$ and $\hat{a}_{k}^\dag$, $\hat{H}_{I}$ describes the interaction between the photonic modes and the dipoles, quantified by the single-dipole light-matter coupling $\chi$, and $\hat{H}_{D}$ is the diamagnetic term which. Assuming the transition to saturate the TRK sum rule \cite{Nataf10}, $D_k=\Omega_0^2/\omega_x$ with $\Omega_0=\sqrt{N} \chi$ the collective light-matter coupling. The diamagnetic term $\hat{H}_D$ can be removed by a Bogoliubov rotation, which leads to the renormalised cavity frequency $ \tilde{\omega}_{k}\equiv\sqrt{ \omega^2_{k} +4 \Omega_{0}^2 } $ and collective light-matter coupling $\tilde{\Omega}_{k}\equiv\Omega_{0}\sqrt{\omega_0/\tilde{\omega}_{k}} $.
The Ising chain, with its periodicity $a$, creates a folding of the photonic dispersion on the border of the first Brillouin zone  
$\omega_{\text{BZ}}=c \pi/a$. A matter mode of the Ising chain indexed by the quasi-momentum $k$ would thus a priori couple with an infinity of photonic modes with wavevector differing by an integer multiple of $\pi/a$. In the following 
we will assume $\omega_{\text{BZ}}-\omega_0\gg\Omega_0$ and neglect Umklapp processes, that is we will assume that the photonic energy on the border of the first Brillouin zone is completely detuned from the dipolar transitions.

\subsection{Polariton spectrum with light-matter interaction within Bose approximation}
We will present and compare the polariton energies calculated solving $\hat{H}_{DI}$ at several orders of approximation.
Initially, we will diagonalise $\hat{H}_{M}$ in terms of bosonic operators $\hat{d}_{O,k}$, where $O \in  [B,\tilde{B} ]$ will indicate either the Bose or the first-order Holstein-Primakoff approximations described in the previous Section, but we will not consider nonlinear terms originating from the light-matter interaction Hamiltonian in \eqref{sigmaI}, effectively treating $\hat{H}_{I}$ in the Bose approximation.
In the next Section we will drop this approximation and expand the first-order Holstein-Primakoff approximation to the full Dicke-Ising Hamiltonian. 

By applying the transformations 
\begin{align}
\hat{b}_{n}=& \frac{1}{\sqrt{N}}\sum_{k} e^{ink}(\alpha_{O,k }\,\hat{d}_{O,k}+\beta_{O,k}\,\hat{d}_{O,-k}^\dag),\nonumber \\
\hat{b}^{\dag}_{n}=& \frac{1}{\sqrt{N}}\sum_{k} e^{-ink}(\beta_{O,k}\,\hat{d}_{O,-k}+\alpha_{O,k }\,\hat{d}_{O,k}^\dag), 
\end{align} 
where $\hat{d}_{O,k}$ and $\hat{d}_{O,k}^\dag$ are the generic bosonic operators which make $\hat{H}_M$ diagonal, and  $\alpha_{O,k }$ and $\beta_{O,k }$ are the generic coefficients of the transformations which can be found in \appref{appA}, the total Hamiltonian $\hat{H}$ can be expressed as:
\begin{align}\label{effcoupling}
\hat{H}=&\sum_{k} \tilde{\omega}_{k} \,\hat{a}_{k}^\dag\hat{a}_{k}+
\sum_{k} E_{O,k} \hat{d}_{O,k}^\dag \hat{d}_{O,k}^{} +E_{O}^{0}+\nonumber \\
&\sum_{k} \Lambda_{O,k} \,(\hat{a}_{k}-\hat{a}_{k}^\dag )( \hat{d}_{O,k}^{}-\hat{d}_{O,-k}^\dag).
\end{align}
In \eqref{effcoupling} $E_{O,k}$ is the single-particle energy, $E_{O}^{0}$ the matter ground-state energy, and 
\be 
\label{Lambda}
\Lambda_{O,k}\equiv\Omega_k F_{O,k},
\ee
is the effective light-matter coupling, with $ F_{O,k}\equiv\alpha_{O,k}-\beta_{O,k}$. The function $ F_{O,k}$, which depends on $\eta$, represents the contribute of the Ising interaction to the effective light-matter coupling $\Lambda_{O,k}$.

We now diagonalise the Hamiltonian $\hat{H} $ by performing the Hopfield-Bogoliubov transformations, described in \appref{appC}, which relates the bosonic operators $\hat{a}_{k}$ and $\hat{d}_{k}$ to the lower and upper polariton operators $\hat{p}_{-,k}^{}$ and $\hat{p}_{+,k}^{}$. The Dicke-Ising Hamiltonian can thus be put in the diagonal form
\be \label{diag}
\hat{H}=\sum_{k} \left ( E_{O,k}^{-} \hat{p}_{-,k}^{\dag}\,  \hat{p}_{-,k}^{} + E_{O,k}^{+} \hat{p}_{+,k}^{\dag}\,  \hat{p}_{+,k}^{} \right )+ E^{G}_O ,
\ee
where 
\be
E^{G}_O=\sum_k \left( E_{O,k}^{-} + E_{O,k}^{+} -\tilde{\omega}_k - E_{O,k} \right)+E_{O}^{0},
\ee
is the ground-state energy of the coupled light-matter system.
The polariton energies $E_{O,k}^{-}$ and $E_{O,k}^{+}$ given by the expression
\be\label{up,lp}
E_{O,k}^{\pm}=\frac{1}{\sqrt{2}}\Bigl[\tilde{\omega}_{k}^2 + E_{O,k}^{2}\pm \Delta_{k}\Bigr]^{\frac{1}{2}},  
\ee
with $ \Delta_{k}=\Bigl[(\tilde{\omega}_{k}^2 - E_{O,k}^{2})^{2} +16 \Lambda_{O,k}^2\, \tilde{\omega}_{k}\, E_{O,k} \Bigr]^{\frac{1}{2}}$, are the eigenvalues of the Hopfield matrix which describes the Hamiltonian. Further details can be found in \appref{appC}.
Diagonalising the matter term within either of the two approximations considered leads to different polariton energies, which, in light of the results obtained for the Ising chain, will differ for level of accuracy.
In the following, we specialise \eqref{up,lp} to the two cases of the Ising Hamiltonian within Bose approximation
($O=B$) and within the first-order Holstein-Primakoff approximation 
($O=\tilde{B}$). Without loss of generality, we will consider $\omega_k>\omega_0$. 

\subsubsection*{\bf{Case} $\mathbold{O=B}$} 
In this case the single-particle energy $E_{B,k}$ is given by \eqref{26} and the expressions of coefficients $\alpha_{B,k}$ and $\beta_{B,k}$ can be found in \appref{appA2}.
The lower and upper polariton energies in \eqref{up,lp}, expanded up to the second order with respect to the normalised Ising coupling $\eta$ and the collective light-matter coupling $\nu\equiv\Omega_0/\omega_0$ take then the form 
\begin{align}
\frac{E^{+}_{B,k}}{\omega_0}\approx&\, \frac{\omega_k}{\omega_0} +  \frac{2\omega_0{\omega}_k \nu^2}{{\omega}_k^2-\omega_0^2},\label{73}\\
\frac{E^{-}_{B,k}}{\omega_0}\approx&\, 1+ 2 \eta \cos{k}  - 2 \eta^2 \cos{k}^2 - \frac{2\omega_0^2 \nu^2}{{\omega}_k^2-\omega_0^2}.\nonumber
\end{align}

\subsubsection*{\bf{Case} $\mathbold{O=\tilde{B}}$ }
We now consider the case in which the matter Hamiltonian is diagonalised within the first-order Holstein-Primakoff approximation, so that $E_{O,k}=E_{\tilde{B},k}$ with $E_{\tilde{B},k}$ from Eq. (\ref{HP1en}) and the coefficients of the transformations are $\alpha_{\tilde{B},k}$ and $\beta_{\tilde{B},k}$, given in \eqref{42}.
The second-order expansions with respect to the normalised coupling strengths $\eta$ and $\nu$ of the polariton energies are thus given by the expressions
\begin{align}\label{bos1polaritons}
\frac{E^{+}_{\tilde{B},k}}{\omega_0}\approx&\, \frac{\omega_k}{\omega_0} +  \frac{2\omega_0{\omega}_k \nu^2}{{\omega}_k^2-\omega_0^2},\\
\frac{E^{-}_{\tilde{B},k}}{\omega_0}\approx&\, 1+ 2 \eta \cos{k}  + 2 \eta^2 \sin{k}^2 - \frac{2\omega_0^2 \nu^2}{{\omega}_k^2-\omega_0^2}.
\nonumber
\end{align}

\subsection{Quantum phase transitions and saturation effect}
The Dicke Hamiltonian including the diamagnetic term presents a superradiant quantum phase transition if 
$D_k<\frac{\Omega_0^2}{\omega_x}$, which is exactly the parameter range excluded by the TRK sum rule 
\cite{Nataf10,Bamba14}. As the Ising coupling modifies both the matter modes energies and their coupling to the photonic field,  it is reasonable to wonder whether the transformed Hamiltonian in \eqref{effcoupling} still maintains this feature. 
A quantum phase transition cannot happen unless the frequency of one of the polariton eigenfrequencies vanishes. 
By requiring that the lower polariton energy $E^{-}_{O,k}$ is not vanishing, we obtain the condition on the effective light-matter coupling $\Lambda_{O,k}$ 
\be\label{condlambda}
\Lambda_{O,k}\ne\pm\frac{\sqrt{\tilde{\omega}_kE_{O,k}}}{2},
\ee
which leads to the inequality
\be\label{condF}
F_{O,k}\leq \sqrt{\frac{E_{O,k}}{\omega_0}},
\ee
always satisfied both for the Bose and the first order Holstein-Primakoff approximations. This demonstrates that the no-go theorems regarding the absence of phase transitions in the Dicke model extend also the the Dicke-Ising model.

It is interesting to notice that the inequality in \eqref{condF} is saturated only for the full Bose approximation, that is for $F_{O,k}=F_{B,k}$ , whereas using the first order Holstein-Primakoff approximation,  $F_{\tilde{B},k}$ is smaller than $F_{B,k}$ of a finite quantity $\eta^2/2$. Since, in general, the effective light-matter coupling is defined as $\Lambda_{O,k}=\tilde{\Omega}_k F_{O,k}$, and the average number of virtual matter excitations in the ground-state of the Ising Hamiltonian 
 to the second order in the normalised dipole-dipole coupling is $\eta^2/2$, we find that within a second-order expansion in $\eta$,
\begin{align}\label{63}
\Lambda_{\tilde{B},k}=\Lambda_{B,k}\left (1-\mathcal{N} \right),
\end{align}
where $\mathcal{N}$ is defined in \eqref{28}.  
The reduction of the effective light-matter coupling in the first order Holstein-Primakoff approximation can in this case be interpreted as a saturation of the transition due to the presence of ground-state virtual excitation, an example of the vacuum nonlinear effects which have recently been investigated in different systems \cite{Garziano14,Frisk17}.  
\begin{figure}[!h]
	\centering
	\includegraphics[scale=0.40]{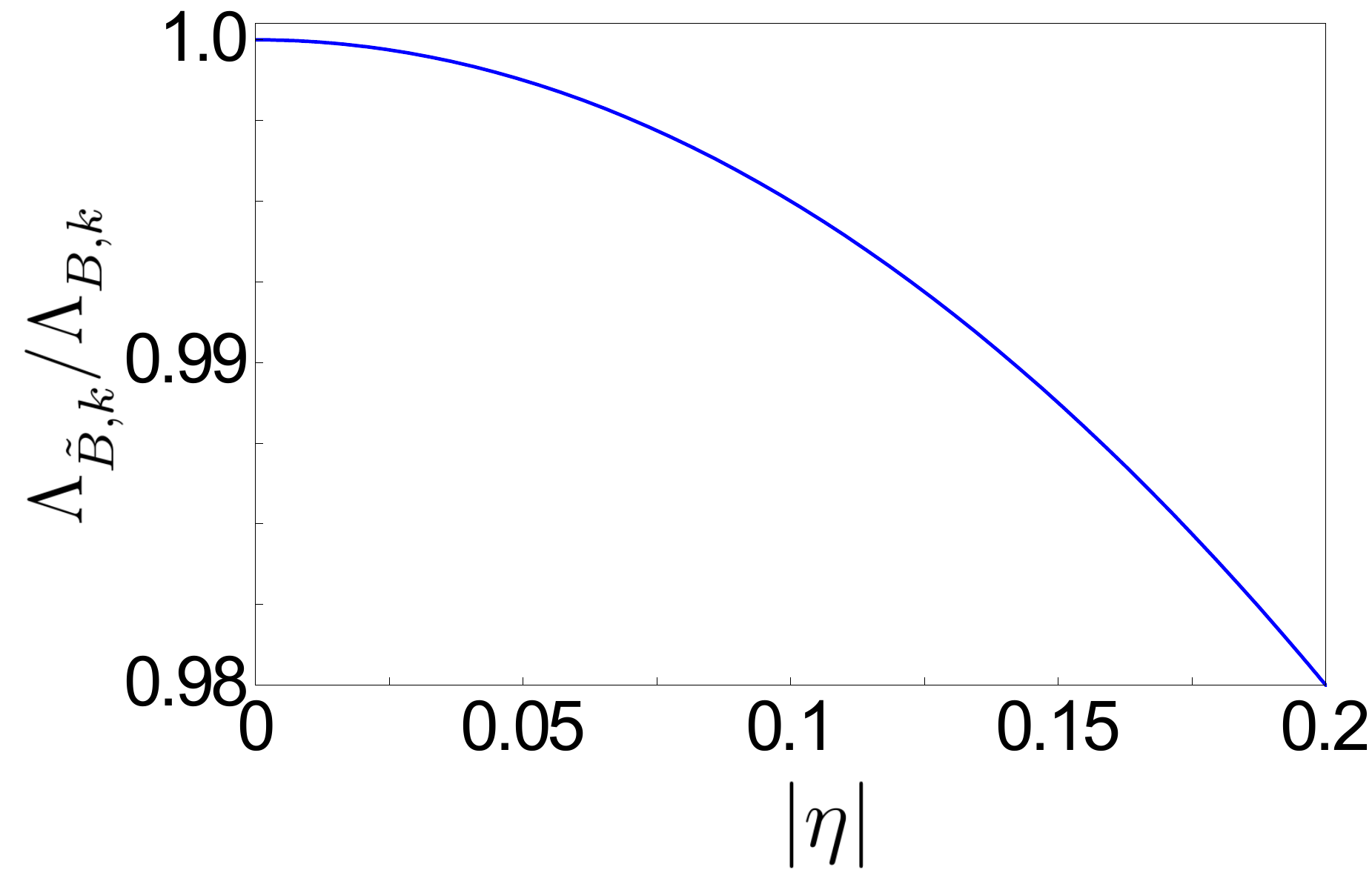}
	\caption{ The effective coupling $\Lambda_{\tilde{B},k}$ normalised with respect to the full Bose $\Lambda_{B,k}$ one as a function of the absolute value of $\eta$. }
	\label{fig:6}
\end{figure}
In \figref{fig:6} we show the behaviour of the effective light-matter coupling $\Lambda_{\tilde{B},k}$  as a function of $\eta$. Within the range $0\le|\eta|\le0.2$, the presence of the virtual photons cause a small but clear reduction of $\sim 2 \% $ respect to the value predicted by the full Bose approximation $\Lambda_{B,k}$. 

\subsection{First-order Holstein-Primakoff approximation of the full Dicke-Ising Hamiltonian}
In the previous Section we considered the light-matter interaction Hamiltonian $\hat{H}_{I}$ within the Bose approximation, neglecting the nonlinear terms deriving from using \eqref{30} in \eqref{sigmaI}.
As in the case of the isolated Ising chain we proved that going beyond the Bose approximation gives relevant improvements to the exciton spectrum, in order to verify the accuracy of the polariton energies in \eqref{bos1polaritons} for large values of the normalised coupling strengths $\eta$ and $\nu$, we proceed including in $\hat{H}_{I}$ the fourth-order nonlinear terms arising from the first Holstein-Primakoff approximation. 
The procedure, analogous to the one of Sec. \ref{secA3} and detailed in \appref{appC}, leads to 
a light-matter interaction term $\hat{H}_I$ of the form
\begin{align}
\hat{H}_I=\hat{H}_I^{(0)}+\hat{H}^{(1)}_{I},
\end{align}
with
\be
\hat{H}_I^{(0)}=\chi \sqrt{\frac{\omega_0}{\tilde{\omega}_k}} \sum_n \sum_k(\hat{a}_k e^{ikn}-\hat{a}_k^\dag e^{-ikn}) (\hat{b}_n-\hat{b}_n^\dag),
\ee
and
\be
\label{HINL}
\hat{H}^{(1)}_{I}=-\frac{\chi}{2}\sqrt{\frac{\omega_0}{\tilde{\omega}_k}}\sum_n \sum_k(\hat{a}_k e^{ikn}-\hat{a}_k^\dag e^{-ikn}) \hat{b}_n^\dag (\hat{b}_n-\hat{b}_n^\dag)b_n.
\ee
The new polariton energies up to the second order in the normalised couplings $\eta$ and $\nu$ can then be calculated as 
\begin{align}\label{upolBtilde}
\frac{E^{+}_{\check{B},k}}{\omega_0}\approx&\,\frac{\omega_k}{\omega_0} +  \frac{2\omega_0 \omega_k \nu^2}{{\omega}_k^2-\omega_0^2},\\ 
\frac{E^{-}_{\check{B},k}}{\omega_0}\approx&\,1+ 2 \eta \cos{k}  + 2 \eta^2 \sin{k}^2 -\nu^2\Biggl ( \frac{2 \omega_0^2}{{\omega}_k^2-\omega_0^2}-\mathcal{F}(N) \Biggr ).\nonumber
\end{align}
Comparing \eqref{upolBtilde} to \eqref{bos1polaritons} we see that the effect of the nonlinear light-matter interaction Hamiltonian $\hat{H}^{(1)}_{I}$ is contained in the term
\be
\label{F}
\mathcal{F}(N)\approx\,\sum_{k'}\frac{1}{N}\frac{\omega_0^2}{2{\omega}_{k'}(1+{\omega}_{k'})}.
\ee
In the large $N$ limit we can substitute the sum in \eqref{F} for an integral, obtaining
\be
\mathcal{F}(N)=\frac{\omega_0 L}{2 c \pi N} \log{\frac{N\left (\omega_0+\frac{c\pi}{L}\right)}{\omega_0+N\frac{c\pi}{L} }},
\ee
which vanishes in the thermodynamic limit $N\rightarrow\infty$ with finite $L$. That is to say, to the considered order, nonlinear terms coming from the light-matter interaction can safely be neglected in the thermodynamic limit.
In the case of a finite number $N$ of dipoles, $\mathcal{F}(N)$ can assume a finite value, giving a non-negligible contribution to the lower polariton energy. By defining the cavity dispersion slope as $\delta\equiv\omega_{BZ}/\omega_0$ such a function  can be written in terms of the new parameter as
\be
\mathcal{F}(N)=\frac{1}{2 \delta} \log{\frac{N+\delta}{1+\delta }},
\ee 
from which we see that also for finite values of $N$, $\mathcal{F}(N)$ vanishes in the non-retarded regime $\delta\rightarrow\infty$.

\subsection{Quantitative comparison}
\begin{figure}[!h]
	\centering
	\includegraphics[scale=0.25
	]{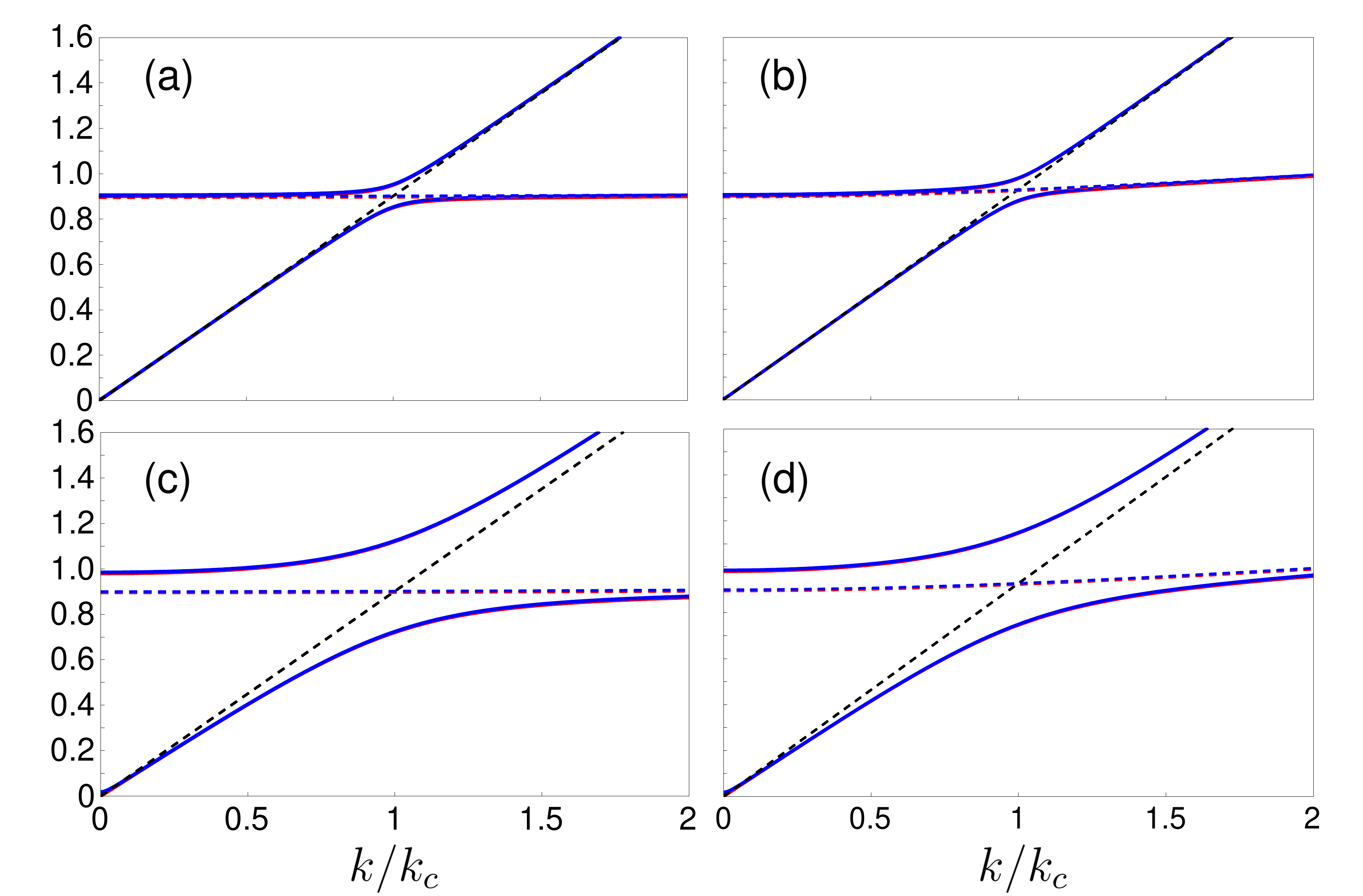}
	\caption{(Color online) Lower and upper polaritons obtained by full Bose treatment (red solid line) and first-order Holstein-Primakoff approximation (blue solid line) for $\eta=-0.05$, as a function of the normalised wavevector $k/k_c$, with $k_c$ the crossing point coordinate. The linear cavity dispersion (black dashed line) and the exciton energies for both the Bose (red dashed line) and first-order  Holstein-Primakoff (blue dashed line) approximation are shown. The plots display cavity dispersion slopes  $\delta=4$ in panels (a) and (c), and $\delta=16$ in panels (b) and (d), and normalised light-matter coupling $\nu=0.05$ in (a) and (b), and $\nu=0.2$ in (c) and (d)}.
	\label{fig:4}
\end{figure}
\begin{figure}[!h]
	\centering
	\includegraphics[scale=0.25
	]{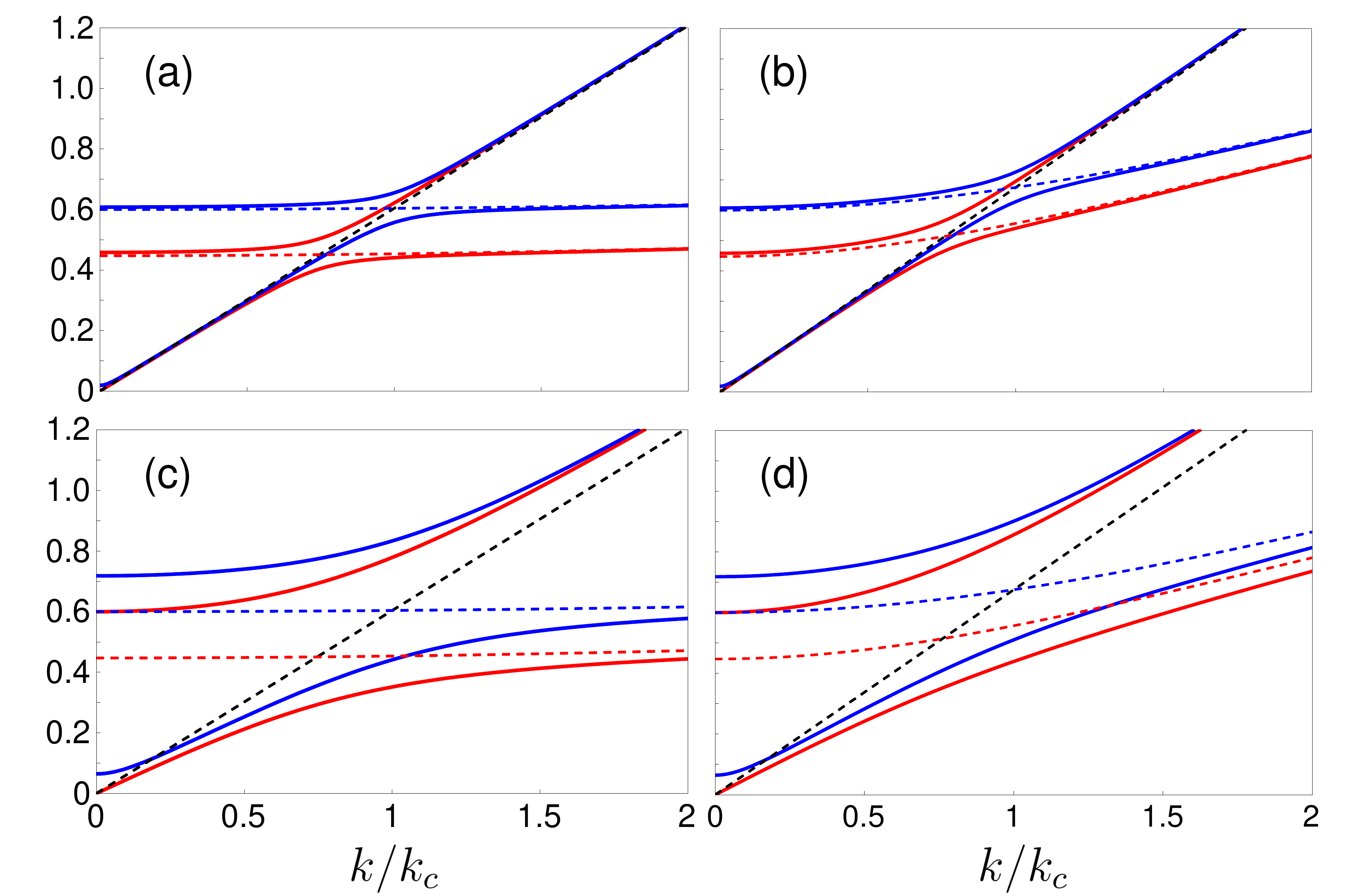}
	\caption{(Color online) Lower and upper polaritons obtained by full Bose treatment (red solid line) and first-order Holstein-Primakoff approximation (blue solid line) for $\eta=-0.2$, as a function of the normalised wavevector $k/k_c$, with $k_c$ the crossing point coordinate. Parameters are the same as in Fig.(\ref{fig:4}).}
	\label{fig:5}
\end{figure}
Figures~\ref{fig:4} and \ref{fig:5} show the dispersive behaviour in the thermodynamic limit of the lower and upper polaritons for both full Bose, $E^{\pm}_{B,k}$ (red line), and first-order Holstein-Primakoff approximation, $E^{\pm}_{\tilde{B},k}$ (blue line), for different Ising and light-matter coupling strengths, and for different slopes of the cavity linear dispersion $\delta$ (black dashed line). Figure~\ref{fig:4} refers to a small normalised Ising coupling $\eta=-0.05$ and both the top panels (a) and (b) correspond to normalised light-matter coupling $\nu=0.05$. The case of ultrastrong light-matter regime is shown in the bottom panels (c) and (d), in which $\nu=0.2$, displaying a sensitive increasing of the splitting between the two polariton energies. The left side panels (a) and (c) differ from the right side panels (b) and (d) for the slope of the cavity frequency. The left panels display a dispersion with  $\delta=4$, while the right panels correspond to  $\delta=16$. In Fig.~\ref{fig:4} we easily observe that the energies of the different approximations perfectly overlap for such low value of the Ising coupling. In \figref{fig:5} a remarkable shift is instead present, with same parameters as Fig.~\ref{fig:4} but a value of the Ising interaction $\eta=-0.2$.
In the thermodynamic limit the main effect of going beyond the Bose approximation, at least at the second order in the coupling strengths, is thus a shift between the exciton energies $E_{\tilde{B},k}$ (blue dashed line) and $E_{B,k}$ (red dashed line).

\section{Conclusions}
We introduced an approach to diagonalise the Ising Hamiltonian in transverse field in terms of bosonic excitations for arbitrary values of the Ising coupling, until the onset of the ferromagnetic phase transition. We then applied such an approach to the Dicke-Ising Hamiltonian, showing how the Ising interaction modifies the usual polaritonic excitations of the Dicke model. 

Our work highlighted a crucial differece between the  Dicke and the Ising models in the regime of large couplings: while in the Dicke model the density of ground-state virtual excitations vanishes in the thermodynamic limit, the same quantity is finite in the Ising case. This ground-state population effectively saturates the transition, causing a reduction of the oscillator strength. Such an example of ground-state nonlinearity, while comparatively small, could be spectroscopically resolved in setups in which the number of dipoles can be controllably modified. 
The model we develop could in particular be implemented in state-of-the-art superconducting circuit QED systems, in which chains of thousand of qubits with engineered interactions have recently been realised. 

We hope our work will not only stimulate further interest in the fields of cavity and circuit QED, but also that the method we introduced will reveal itself useful to investigate a larger class of many body systems in which interactions are strongly enough to modify the nature of the ground state.

\section{Acknowledgement} 
The authors acknowledge support from EPSRC grant EP/M003183/1. 
S.D.L. is a Royal Society Research Fellow.

\appendix

\section{Diagonalisation of the Hamiltonian for the transverse field Ising model.} \label{appA}

\subsection{Exact diagonalisation} \label{appA1}

The Hamiltonian in \eqref{1} can be diagonalised exactly by performing the Jordan-Wigner transformation
\begin{align} \nonumber
\hat{\sigma}_{n}^{-}=\prod_{j<n}(1-2 \,\hat{c}^{\dag}_{j}\hat{c}_{j})\hat{c}_{n},\\ \label{3}
\hat{\sigma}_{n}^{+}=\prod_{j<n}(1-2 \,\hat{c}^{\dag}_{j}\hat{c}_{j})\hat{c}^{\dag}_{n},
\end{align}
where $ \hat{c}^{\dag}_{n} $ and $ \hat{c}_{n} $ are fermionic creation and annihilation operators satisfying the anticommutation rules: 
\be \Bigl\{\hat{c}_{n},\hat{c}_{m}^\dag\Bigr\}=\delta_{nm}\;,\;  \Bigl\{\hat{c}_{n}^\dag,\hat{c}_{m}^\dag\Bigr\}=\Bigl\{\hat{c}_{n},\hat{c}_{m}\Bigr\}=0.
\ee 
The Hamiltonian in \eqref{1} then takes the form 
\begin{align} \nonumber
\hat{H}_{\rm F}=\,& \omega_{\rm 0} \sum_{n=1}^{N}\hat{c}_{n}^\dag \hat{c}_{n}+J\sum_{n=1}^{N-1}(
\hat{c}_{n}^\dag \hat{c}_{n+1}-\hat{c}_{n}\hat{c}_{n+1}^\dag)\\\label{4}
&+J\sum_{n=1}^{N-1}(\hat{c}_{n}^\dag \hat{c}_{n+1}^\dag-\hat{c}_{n}\hat{c}_{n+1}).
\end{align}
We can now pass in momentum space defining the operators $\hat{c}_{k}$ and $\hat{c}_{k}^\dag$  by the relations
\begin{align} \nonumber
\hat{c}_{n}&=\frac{1}{\sqrt{N}}\,e^{-i\pi/4}\sum_{k} e^{ink}\, \hat{c}_{k},\\ \label{6}
\hat{c}^\dag_{n}&=\frac{1}{\sqrt{N}}\,e^{i\pi/4}\sum_{k} e^{-ink}\, \hat{c}^\dag_{k},
\end{align}
where the phase factor $ \exp(-i \pi/4) $ is introduced in order to guarantee real coefficients. The resulting Hamiltonian is then put in the diagonal form in \eqref{7} in terms of the fermionic operators
\begin{align}\nonumber
\dfk&=\alpha_{F,k}\,\hat{c}_{k}+\beta_{F,k}\,\hat{c}^\dag_{-k},\\\label{8}
\dfck&=\beta_{F,k}\,\hat{c}_{-k}+\alpha_{F,k}\,\hat{c}^\dag_{k},
\end{align}  
with 
\begin{align}\nonumber
\alpha_{F,k}&= \frac{\omega_{\rm 0}+2J\cos k +E_{F,k}}{\sqrt{\Bigl(\omega_{\rm 0}+2J\cos k +E_{F,k}\Bigr)^2+4J^2 \sin^2 k}},\\\label{10}
\beta_{F,k}&= -\frac{2J\sin k}{\sqrt{\Bigl(\omega_{\rm 0}+2J\cos k +E_{F,k}\Bigr)^2+4J^2 \sin^2 k}}.
\end{align}

\subsection{The Bose approximation} \label{appA2}

The Hamiltonian in \eqref{1} can be mapped into a fully bosonic one through the Holstein-Primakoff transformation (see \eqref{14}). 
If one retains only the first term of the expansions the Hamiltonian becomes:
\begin{align} \nonumber
\hat{H}_{\rm B}=\,& \omega_{\rm 0} \sum_{n=1}^{N}\hat{b}_{n}^\dag \hat{b}_{n}+J\sum_{n=1}^{N-1}(
\hat{b}_{n}^\dag \hat{b}_{n+1}+\hat{b}_{n}\hat{b}_{n+1}^\dag)\\ \label{18}
&+J\sum_{n=1}^{N-1}(\hat{b}_{n}^\dag \hat{b}_{n+1}^\dag+\hat{b}_{n}\hat{b}_{n+1}).
\end{align}
Following a procedure similar to the one used in the exact diagonalisation of the fermionic case, one can pass in momentum space by the relations
\begin{align} \nonumber
\hat{b}_{n}&=\frac{1}{\sqrt{N}}\sum_{k} e^{ink}\, \hat{b}_{k},\\ \label{19}
\hat{b}^\dag_{n}&=\frac{1}{\sqrt{N}}\sum_{k} e^{-ink}\, \hat{b}^\dag_{k}.
\end{align}
After substituting \eqref{19} in \eqref{18}, the Hamiltonian can be put in the diagonal form in \eqref{25} by the Bogoliubov transformation
\begin{align}\nonumber
\dbk &=\alpha_{B,k}\,\hat{b}_{k}+\beta_{B,k}\,\hat{b}^\dag_{-k},\\\label{21}
\dbck&=\beta_{B,k}\,\hat{b}_{-k}+\alpha_{B,k}\,\hat{b}^\dag_{k},
\end{align}
with
\begin{align}\nonumber
\alpha_{B,k}&= \frac{\omega_{\rm 0}+2J\cos k +E_{B,k}}{\sqrt{\Bigl(\omega_{\rm 0}+2J\cos k +E_{B,k}\Bigr)^2-4J^2 \cos^2 k}}
,\\\label{23}
\beta_{B,k}&= \frac{2J\cos k}{\sqrt{\Bigl(\omega_{\rm 0}+2J\cos k +E_{B,k}\Bigr)^2-4J^2 \cos^2 k}}.
\end{align} 

Introducing the ground state $\ket{G}$, defined by the usual relations $\dbk\ket{G}=0$ and inverting \eqref{21} we can then calculate the ground state population of virtual excitations
\begin{align}
\mathcal{N}&\equiv \bra{G}\hat{b}^{\dag}_{n}\hat{b}_{n}\ket{G}=\frac{1}{N}\sum_k \bra{G}\hat{b}^{\dag}_{k}\hat{b}_{k}\ket{G}\nonumber\\&= \frac{1}{N} \sum_k \rvert \beta_{B,k} \lvert^2= \frac{\eta^2}{2}+O(\eta^3).
\end{align}

\section{Diagonalisation of the bosonic matter Hamiltonian at the first-order Holstein-Primakoff approximation} \label{appB}

After inserting \eqref{34} into \eqref{32}, neglecting the nonlinear normally-ordered terms which vanish in the zero- and one-particle subspace, one obtains the Hamiltonian in \eqref{36}, where the linear coefficients are
\begin{align} \nonumber
\mathcal{A}_k=&\omega_0 (\alpha_{\tilde{B},k }^2+\beta_{\tilde{B},k}^2)+2 (\alpha_{\tilde{B},k }+\beta_{\tilde{B},k})^2J\cos k,\\ \nonumber
\mathcal{B}_k=&\omega_0\, \alpha_{\tilde{B},k }\,\beta_{\tilde{B},k}+2 (\alpha_{\tilde{B},k }+\beta_{\tilde{B},k})^2J\cos k,\\
\mathcal{C}_k=&\omega_0\, \beta_{\tilde{B},k}^2+J\cos k (2\alpha_{\tilde{B},k }\beta_{\tilde{B},k}+\beta_{\tilde{B},k}^2),
\end{align}
and the nonlinear ones are
\begin{widetext}
\begin{align} 
f(k,k')=&2\left(\alpha_{\tilde{B},k } \beta_{\tilde{B},k'}+\alpha_{\tilde{B},k' } \beta_{\tilde{B},k}\right)\left(\alpha_{\tilde{B},k}+\beta_{\tilde{B},k}\right)\left(\alpha_{\tilde{B},k'}+\beta_{\tilde{B},k' }\right)\left( \cos{k}+\cos{k'}\right)\nonumber \\
&+2\,\alpha_{\tilde{B},k }\beta_{\tilde{B},k } \cos{k'}\left(\alpha_{\tilde{B},k'}+\beta_{\tilde{B},k' }\right)^2+\nonumber 2\,\beta_{\tilde{B},k'}^2 \cos{k}\left(\alpha_{\tilde{B},k}+\beta_{\tilde{B},k }\right)^2,
&\\\nonumber
g(k,k')=&4\, \beta_{\tilde{B},k'}\left(\alpha_{\tilde{B},k }+\beta_{\tilde{B},k}\right )^2 \left(\alpha_{\tilde{B},k'}+\beta_{\tilde{B},k' }\right)\left( \cos{k}+\cos{k'}\right)+
2\,\alpha_{\tilde{B},k }^2 \cos{k'}\left(\alpha_{\tilde{B},k'}+\beta_{\tilde{B},k' }\right)^2\\&+4\,\beta_{\tilde{B},k'}^2 \cos{k}\left(\alpha_{\tilde{B},k}+\beta_{\tilde{B},k }\right)^2+2\,\beta_{\tilde{B},k}^2 \cos{k'}\left(\alpha_{\tilde{B},k'}+\beta_{\tilde{B},k' }\right)^2,\nonumber \\
h(k,k')=&2\, \beta_{\tilde{B},k} \beta_{\tilde{B},k'} \left( \cos{k}+\cos{k'}\right)\left(\alpha_{\tilde{B},k}+\beta_{\tilde{B},k}\right)\left(\alpha_{\tilde{B},k'}+\beta_{\tilde{B},k' }\right)+
2\,\beta_{\tilde{B},k}^2 \cos{k'}\left(\alpha_{\tilde{B},k'}+\beta_{\tilde{B},k' }\right)^2.
\end{align}
The coefficients $\alpha_{\tilde{B},k }$ and $\beta_{\tilde{B},k }$ can then be determined perturbatively  to the second order through \eqref{S1} by writing
\begin{align}\nonumber
\beta_{\tilde{B},k}&\approx\bk{0}+\bk{1}\,\eta+\bk{2}\,\eta^2,\nonumber\\
\alpha_{\tilde{B},k }&\approx\sqrt{1+\beta_{\tilde{B},k}^2}\approx 1+\frac{\beta_{\tilde{B},k}^{(0)2}}{2}+\bk{0}\bk{1}\eta
+\frac{\Bigl(\beta_{\tilde{B},k}^{(1)2}
	+2\bk{0}\bk{2}\Bigr)}{2}\eta^2,
\end{align}
and solving analytically the resulting algebraic system.
\end{widetext}

\section{Diagonalisation of the Bosonic light-matter Hamiltonian} \label{appC}
The  bosonic light-matter Hamiltonian in \eqref{effcoupling} can be diagonalised using an approach originally due to Hopfield \cite{Hopfield58}. For the sake of clarity in this Appendix we will drop the $O$ subscript from the bosonic operators and the effective coupling as the outlined procedure is valid for arbitrary bosonic operators.
 
In the basis $ \hat{\mathbf{v}}_{k}=\Bigl[ \hat{a}_{k}\,\, \hat{d}_{k} \,\,  \hat{a}_{-k}^\dag \,\,  \hat{d}_{-k}^\dag   \Bigr]^{T}$, the Hopfield matrix for the Hamiltonian of \eqref{effcoupling}  is
\be \label{matrix}
\mathcal{M}_{k}=\begin{pmatrix}
	\tilde{\omega}_{k} && -\Lambda_{k} && 0 && \Lambda_{k} \\\\
	-\Lambda_{k} && E_{k}  && \Lambda_{k} && 0\\ \\
	0 && -\Lambda_{k} && -\tilde{\omega}_{k} && \Lambda_{k}\\ \\
	-\Lambda_{k} && 0  && \Lambda_{k} && -E_{k} 
\end{pmatrix}
\ee
so that
\be 
\hat{H}=\frac{1}{2}\hat{\mathbf{v}}_{k}^\dag \mu\mathcal{M}_{k} \hat{\mathbf{v}}_{k},
\ee
 with  $ \mu=$diag$[1,1,-1,-1] $ the bosonic metric matrix.\\
The eigenvalues of $\mathcal{M}_{k}$ are
\be
E_{k}^{\pm}=\frac{1}{\sqrt{2}}\Bigl[\tilde{\omega}_{k}^2 + E_{k}^{2}\pm \Delta_{k}\Bigr]^{\frac{1}{2}},  
\ee
with $ \Delta_{k}=\Bigl[(\tilde{\omega}_{k}^2 - E_{k}^{2})^{2} +16 \Lambda_{k}^2\, \tilde{\omega}_{k}\, E_{k} \Bigr]^{\frac{1}{2}}$.\\
If we define the vector $ \hat{\mathbf{p}}_{k}=\Bigl[ \hat{p}_{-,k}\,\,\,\, \hat{p}_{+,k} \,\,\,\,  \hat{p}_{-,-k}^\dag \,\,\,\,  \hat{p}_{+,-k}^\dag   \Bigr]^{T} $, the Hamiltonian can then be put in the diagonal form
\be 
H=\frac{1}{2}\hat{\mathbf{p}}_{k}^\dag \mu\mathcal{D}_{k} \hat{\mathbf{p}}_{k},
\ee 
where $ \mathcal{D}_{k}=\text{diag}[E_k^{-}, E_k^{+}, -E_k^{-}, -E_k^{+}]$.\\
The vectors $ \hat{\mathbf{p}}_{k}   $ and $ \hat{\mathbf{v}}_{k} $ are related by the relation $ \hat{\mathbf{p}}_{k}=U^\dag \mu \,\hat{\mathbf{v}}_{k} $, where  
$ U $ is the unitary transformation matrix. The extended expressions for the lower- and upper-polariton operators $ \hat{p}_{-,k} $ and  $ \hat{p}_{+,k} $ are then:
\begin{align} \label{p}
\hat{p}_{-,k}&=x_{-} \hat{a}_{k} + y_{-} \hat{d}_{k}+ w_{-}\hat{a}_{-k}^\dag+z_{-}\hat{d}_{-k}^\dag,\nonumber\\
\hat{p}_{+,k}&=x_{+} \hat{a}_{k} + y_{+} \hat{d}_{k}+ w_{+}\hat{a}_{-k}^\dag+z_{+}\hat{d}_{-k}^\dag,
\end{align}
where the coefficients of the transformations, which can always be chosen real, are the components of the eigenvectors:
\begin{widetext}
\be
\begin{pmatrix} \label{eigenvectors}
	x_{\pm}  \\ 
	y_{\pm}\\
	w_{\pm} \\
	z_{\pm}
\end{pmatrix}= \Biggl\{ 4 E_{k}^{\pm}\tilde{\omega}_{k}  \Biggl[1+\frac{4 \Lambda_{k}^{2} E_{k}\, \tilde{\omega}_{k}\,  }{(E_{k}^{\pm 2}-E_{k}^2 )^2}\Biggr]  \Biggr\}^{-\frac{1}{2}}\begin{pmatrix} 
	\tilde{\omega}_{k}+E_{k}^{\pm}  \\ 
	\frac{2 \Lambda_{k} \tilde{\omega}_{k}}{E_{k}^{\pm}-E_{k}}\\
	-(\tilde{\omega}_{k}-E_{k}^{\pm}) \\
	\frac{2 \Lambda_{k} \tilde{\omega}_{k}}{E_{k}^{\pm}+E_{k}},
\end{pmatrix}.
\ee
\end{widetext}
where normalisation factor is found by imposing that the operators  $ \hat{p}_{-,k} $ and  $ \hat{p}_{+,k} $ to obey bosonic commutation rules $\left [\hat{p}_{\pm,k}, \hat{p}_{\pm,k'}^\dag \right]=\delta_{k,k'}$.
The inverse relation $ \hat{\mathbf{v}}_{k}=U \mu \,\hat{\mathbf{p}}_{k} $ allows us to express the operators $\hat{a}_{k}  $ and $ \hat{d}_{k} $ as a linear combinations of $ \hat{p}_{-,k} $ and  $ \hat{p}_{+,k} $ as
\begin{align}
\hat{a}_{k}&=x_{-} \hat{p}_{-,k} + x_{+} \hat{p}_{+,k}- w_{-}\hat{p}_{-,-k}^{\dag}-w_{+}\hat{p}_{+,-k}^{\dag},\nonumber \\\label{67}
\hat{d}_{k}&=y_{-} \hat{p}_{-,k} + y_{+} \hat{p}_{+,k}- z_{-}\hat{p}_{-,-k}^{\dag}-z_{+}\hat{p}_{+,-k}^{\dag}.
\end{align}

\section{Diagonalization of the light-matter Hamiltonian at the first order Holstein-Primakoff approximation}\label{B}
Analogously to what has been done for the isolated Ising chain, we aim to diagonalise the full Dicke-Ising Hamiltonian in the first order Holstein-Primakoff approximation, obtained by plugging \eqref{30} in \eqref{H}, including the nonlinear part of the light-matter Hamiltonian in \eqref{HINL}.
The multimode equivalent of \eqref{34} reads
\begin{align}\label{66}
\hat{a}_{k}&=x_{\check{B},-}\, \hat{p}_{-,k} + x_{\check{B},+}\, \hat{p}_{+,k}- w_{\check{B},-}\,\hat{p}_{-,-k}^{\dag}-w_{\tilde{B},+}\,\hat{p}_{+,-k}^{\dag},\nonumber \\
\hat{b}_{k}&=y_{\check{B},-} \,\hat{p}_{-,k} + y_{\check{B},+}\, \hat{p}_{+,k}- z_{\check{B},-}\,\hat{p}_{-,-k}^{\dag}-z_{\check{B},+}\,\hat{p}_{+,-k}^{\dag},
\end{align}
where the Hopfield coefficients will be determined by diagonalising the quadratic part of the normally-ordered Hamiltonian expressed in terms of the polaritonic operators.
By requiring that all the non-diagonal terms must go to zero, we then recover a nonlinear system of equations for the Hopfield coefficients. In order to simplify the analytic calculation, we first expand the coefficients $x_{\check{B},\pm}$,$y_{\check{B},\pm}$, $z_{\check{B},\pm}$, and $w_{\check{B},\pm}$ respect to the normalised couplings $\eta$ and $\nu$ up to the second order
\begin{widetext}
\begin{align}
\nonumber 
x_{\check{B},\pm}\approx&  \,x_{\check{B},\pm}^{(0)}+x_{\check{B},\pm}^{(1,\eta)} \eta +x_{\check{B},\pm}^{(1,\nu)} \nu+x_{\check{B},\pm}^{(2,\eta)} \eta^2 +x_{\check{B},\pm}^{(2,\eta\nu)}\eta \nu+ x_{\check{B},\pm}^{(2,\nu)} \nu^2,\\
\nonumber  y_{\check{B},\pm}\approx& \,y_{\check{B},\pm}^{(0)}+y_{\check{B},\pm}^{(1,\eta)} \eta +y_{\check{B},\pm}^{(1,\nu)} \nu+y_{\check{B},\pm}^{(2,\eta)} \eta^2 +y_{\check{B},\pm}^{(2,\eta\nu)}\eta \nu+y_{\check{B},\pm}^{(2,\nu)} \nu^2,\\
\nonumber  z_{\check{B},\pm}\approx& \,z_{\check{B},\pm}^{(0)}+z_{\check{B},\pm}^{(1,\eta)} \eta +z_{\check{B},\pm}^{(1,\nu)} \nu+z_{\check{B},\pm}^{(2,\eta)} \eta^2 +z_{\check{B},\pm}^{(2,\eta\nu)}\eta \nu+z_{\check{B},\pm}^{(2,\nu)} \nu^2,\\
w_{\check{B},\pm}\approx& \,w_{\check{B},\pm}^{(0)}+w_{\check{B},\pm}^{(1,\eta)} \eta +w_{\check{B},\pm}^{(1,\nu)} \nu+z_{\check{B},\pm}^{(2,\eta)} \eta^2 +w_{\check{B},\pm}^{(2,\eta\nu)}\eta \nu+w_{\check{B},\pm}^{(2,\nu)} \nu^2.
\end{align}
By solving the system of equations, we obtain the expressions of the coefficients:
\begin{align}
\nonumber x_{\check{B},-}\approx&\,1-\frac{2 \omega_0^4\nu^2}{(\omega_0^2-{\omega}_k^2)^2},\\   \nonumber
x_{\check{B},+}\approx&\,\frac{\omega_0^{3/2}\nu}{\sqrt{{\omega}_k}({\omega}_k-\omega_0)}+\frac{\cos{k}(\omega_0+{\omega}_k) \omega_0^{3/2} \eta \nu}{\sqrt{{\omega}_k}({\omega}_k-\omega_0)^2},\\
\nonumber y_{\check{B},-}\approx&\,\frac{\omega_0^{3/2}\nu}{\sqrt{{\omega}_k}(\omega_0-{\omega}_k)}-\frac{4 \cos{k}\sqrt{{\omega}_k}\omega_0^{5/2}\eta \nu}{({\omega}_k-\omega_0)^2(\omega_k+\omega_0)},\\   \nonumber
y_{\check{B},+}\approx&\,1+\frac{1}{2}\eta^2 \cos{k}^2-\frac{2\omega_0^4 \nu^2}{({\omega}_k-\omega_0)^2({\omega}_k+\omega_0)^2},\\
\nonumber 
z_{\check{B},-}\approx&\,\frac{\omega_0^{3/2}\nu}{\sqrt{{\omega}_k}({\omega}_k+\omega_0)}-\frac{4 \cos{k} \sqrt{{\omega}_k}\omega_0^{5/2}\eta \nu}{({\omega}_k^2-\omega_0^2)({\omega}_k+\omega_0)}, \\  
z_{\check{B},+}\approx&\nonumber \,\eta \cos{k}-\eta^2\left (2 \cos{k}^2-\frac{1}{2}\right) -\nu^2\Biggl(\frac{\omega_0^2}{\omega_0^2-{\omega}_k^2}+\sum_{k'}\frac{1}{N}\frac{\omega_0^2}{2({\omega}_{k'}\omega_0+{\omega}_{k'}^2)}\Biggr),\\
\nonumber
w_{\check{B},-}\approx&\,-\frac{\omega_0^4\nu^2}{{\omega}_k^2({\omega}_k^2-\omega_0^2)},\\
 w_{\check{B},+}\approx&\,\frac{\omega_0^{3/2}\nu}{\sqrt{{\omega}_k}({\omega}_k+\omega_0)}+\frac{\cos{k}({\omega}_k-\omega_0)^3 \eta \nu}{\sqrt{{\omega}_k\omega_0}({\omega}_k^2-\omega_0^2)}.
\end{align} 
\end{widetext}

\bibliography{ErikaLuigibib}

\end{document}